\documentclass[epjST]{svjour}
\usepackage{graphics}
\usepackage{amssymb,euscript,latexsym,amsmath}
\usepackage[dvips]{epsfig}
\usepackage{color}
\usepackage{bm}
\usepackage{enumerate}
\usepackage{hyperref}
\usepackage{setspace}
\usepackage{subfigure} 
\usepackage{float}
\usepackage{xspace}

\newcommand{\poincare}{Poincar\'e~}

\begin{document}
\title{Periodic and Chaotic Flapping of Insectile Wings}
\author{Yangyang Huang\and Eva Kanso
\thanks{\emph{Corresponding author: kanso@usc.edu} }%
}                     
%
%
\institute{Aerospace and Mechanical Engineering, University of Southern California, Los Angeles, CA 90089}
\date{Received: April 24, 2015}
%
\abstract{
Insects use flight muscles attached at the base of the wings to produce impressive wing flapping frequencies. 
The maximum power output of these flight muscles is insufficient to maintain such wing oscillations  
unless there is good elastic storage of energy in the insect flight system. Here, we explore the intrinsic self-oscillatory behavior of an insectile wing model, consisting of two rigid wings connected at their base by an elastic torsional spring.  We study the wings behavior  as a function of the total energy and spring stiffness.  Three types of behavior are identified: end-over-end rotation, chaotic motion, and periodic flapping. Interestingly, the region of periodic flapping decreases as energy increases but is favored as stiffness increases. These findings are consistent with the fact that insect wings and flight muscles  are stiff. They further imply that, by adjusting their muscle stiffness to the desired energy level, insects can maintain periodic flapping mechanically for a range of operating conditions.
} 
\maketitle
\section{Introduction}
\label{intro}

The flapping of insect wings is a marvelous example of autonomous oscillations. 
Insects use flight muscles attached at the base of the wings~\cite{Pringle2003} to produce wing flapping frequencies that, 
in certain species, far exceed the animals' neural capacity \cite{Mayr2012}. In these species, the contractile activity of flight 
muscles is maintained by a self-oscillatory mechanism that is under mechanical, not nervous, control \cite{Ellington1985}. 
Calculations of mechanical power suggest that the maximum power output of flight muscles is adequate for the aerodynamic power requirements, but insufficient to also oscillate the wings' mass unless there is good elastic storage of the inertial energy \cite{Ellington1985}. Elastic energy could be stored in several components of the insect flight system, including flight muscles. Insect flight muscles are very stiff such that, even at small operating strains, they can store elastically much of the inertial energy of the oscillating wing \cite{Ellington1985}.

\begin{figure}[!t]
\centering
\includegraphics[width=0.35\linewidth]{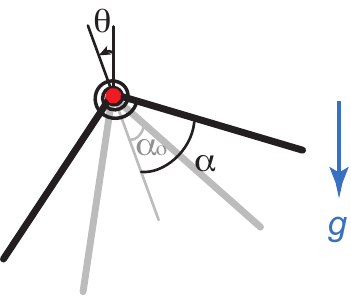}
\caption{\footnotesize Insectile wing model: two rigid wings connected by a torsional spring of stiffness $\kappa$. The wings are free to flap (angle $\alpha$) and rotate about their base point (angle $\theta$).}
\label{fig:wings}
\end{figure}

In this paper, we formulate and analyze the dynamics of an idealized insectile wing model, consisting of two rigid wings connected at their base by an elastic torsional spring (see Fig.~\ref{fig:wings}). Our goal is to examine the instrinsic self-oscillations of the wings under their own inertial load by drawing upon tools from nonlinear dynamical systems. This approach is aimed at complementing past and ongoing research efforts on flapping insect wings. 
The impressive aerodynamic performance of insects have stimulated a great deal of interest among biologists, physicists and engineers, whose focus 
is to decipher the biomechanics underlying  insect flight and to translate this knowledge into design principles for engineered micro-air vehicles.
Here, we give a brief overview  of such work, with particular bias towards the aerodynamics of flapping flight. 
Early attempts at explaining the high-lift mechanisms in insect flight use 
quasi-steady aerodynamic analysis \cite{WeisJensen1956}. This approach provides valuable qualitative insights into 
the mechanisms of force generation in flapping flight but underestimates unsteady aerodynamic effects.
The unsteady mechanisms responsible for the generation of remarkable lift forces in flapping flight has been examined in recent experimental \cite{Ellington1996,Dickinson1999,Spedding2003,Birch2003,Thomas2004,Douglas2005} and theoretical \cite{Ramamurti2002,Minotti2002,Sun2002,Sun2004,Wang2000PRL,Wang2000JFM,Wang2004,Wang2007} studies, mostly emphasizing the importance of leading-edge and wake vorticity in force production \cite{Sane2003,Wang2005}. Attention turned recently to the stability of flapping flight in response to environmental disturbances \cite{Sun2014}, {with}
conflicting accounts of intrinsic instability \cite{Sun2005,Sun2007} and passive stability \cite{Hedrick2009,Faruque2010,Cheng2011}. Given that 
an assessment of the passive stability of live organisms is not feasible experimentally, proxy models of inanimate flyers are proposed in~\cite{Childress2006,Weathers2010} and their passive stability is discussed in \cite{Liu2012,Huang2014}.

The present paper does not  focus on wing aerodynamics, but on the intrinsic nonlinear dynamics of insect-like wings when subject to their own weight and elastic energy storage.
Evolution and natural selection have made  insect wings stiff enough to withstand the aerodynamic load \cite{Rees1975}. Unlike birds and bats which have active muscles in their modified forelimbs to control the wing shape, insect wings twist and camber due to their elastic properties \cite{Wootton1992}. Wing flexural stiffness  varies along the wing \cite{Combes2003I,Combes2003II}. However, we consider here a ``lumped" spring model, where the wing and muscle stiffness  are both modeled via an elastic torsional spring at the based of two rigid wings. We analyze the stability of this insectile wing system around the vertically down position and explore its global behavior as a function of total energy and spring stiffness. Interestingly, we identify regions of rotational motion (rolling end over end),  chaotic behavior, and periodic flapping. The region of periodic flapping decreases as energy increases, and increases as spring stiffness increases. We conclude by discussing the potential implications of these findings on flapping flight. 
%
%

\section{Model}
\label{sec:model}

\begin{figure*}[!t]
\centering
\includegraphics[width=\linewidth]{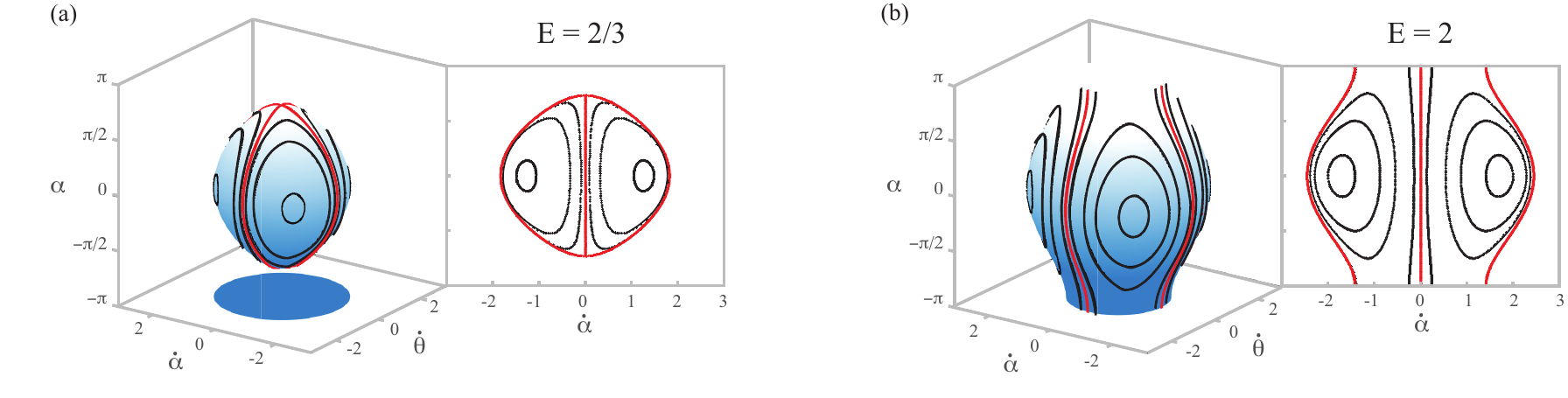}
\caption{\footnotesize Uncoupled wings ($\kappa = 0$): Poincar\'{e} sections at $\theta = 0$ for (a) $E=E_1 + E_2 = 2/3$ and (b) $E=2$. Insets show projections  of the Poincar\'{e} sections onto the plane $\dot{\theta}=0$. As energy increases, the compact energy surface  becomes periodic in $\alpha$. 
}
\label{fig:zerostiff}
\end{figure*}

We consider two rigid wings, of mass $m$ and length $l$ each, connected at their base by a torsional spring of stiffness $\kappa$ (see Fig.~\ref{fig:wings}). We let $2\alpha$ denote the opening angle of the wings, or the  ``flapping" degree of freedom, and $2\alpha_o$ denote the spring rest angle such that the internal spring torque is equal to $2\kappa(\alpha-\alpha_o)$. Further, the wings are free to rotate about their base point; their orientation angle from the vertical is denoted by $\theta$. Ignoring aerodynamic forces, and under gravitational effects only, the equations of motion governing the behavior of $\alpha$ and $\theta$ are given by
\begin{equation}
\begin{aligned}
(\frac{2}{3}ml^2)\ddot{\theta} &= -mgl\cos\alpha \sin\theta \\
(\frac{2}{3}ml^2)\ddot{\alpha} &= -mgl\sin\alpha \cos\theta - 4\kappa(\alpha-\alpha_o).
\end{aligned}
\end{equation}
Here, $g$ is the gravitational constant. In order to re-write the above equations in non-dimensional form, we use the  length scale $l^*=l$ and time scale $t^*=\sqrt{2l/3g}$, and introduce the dimensionless spring constant $\kappa^\ast = 4 \kappa/mgl$. To this end, we get
\begin{equation}
\begin{split}
\ddot{\theta} &= -\cos\alpha \sin\theta \\
\ddot{\alpha} &= -\sin\alpha \cos\theta - \kappa(\alpha-\alpha_o),
\label{eq:eom}
\end{split}
\end{equation}
where we dropped the $()^\ast$ notation with the understanding that all quantities are now non-dimensional. The total energy $E$ of the wings is conserved,
\begin{equation}
\begin{aligned}
E = \frac{1}{2}(\dot{\alpha}^2+\dot{\theta}^2)-\cos\alpha\cos\theta+\frac{1}{2}\kappa(\alpha-\alpha_o)^2.
\label{eq:energy}
\end{aligned}
\end{equation}

For zero spring stiffness, $\kappa = 0$, this wing model is equivalent to a system of two uncoupled rigid pendula. This system is integrable because it admits two conserved quantities, namely, $E_1 = \frac{1}{2}(\frac{1}{2}(\dot{\alpha}+\dot{\theta})^2-\cos(\alpha+\theta))$ and $E_2 = \frac{1}{2}(\frac{1}{2}(\dot{\alpha}-\dot{\theta})^2-\cos(\alpha-\theta))$, such that $ E = E_1 + E_2$.  Poincar\'{e} section  in the four-dimensional phase space $(\theta,\alpha,\dot{\theta},\dot{\alpha})$ is taken at $\theta = 0$. A depiction is shown in Fig.~\ref{fig:zerostiff} for two total energy levels $E =2/3$ and $E = 2$. At lower $E$, the energy surface is compact and closed, while at higher $E$, the energy surface is periodic in the $\alpha$-direction.

For infinite spring stiffness, $\kappa = \infty$, the two wings behave as one rigid pendulum, whose dynamics is governed by $\ddot{\theta} = -\cos\alpha \sin\theta$. The frequency of oscillations of the pendulum depends on its shape, defined by the fixed parameter $\alpha$. As $\alpha$ increases from $0$ to $\pi$, the oscillation frequency first decreases then increases. It reaches zero at $\alpha = \pi/2$, where the gravitational torques on the two wings of the rigid pendulum are perfectly balanced and the system is in a state of equilibrium for all $\theta$.

\begin{figure*}
\centering
\includegraphics[width=\linewidth]{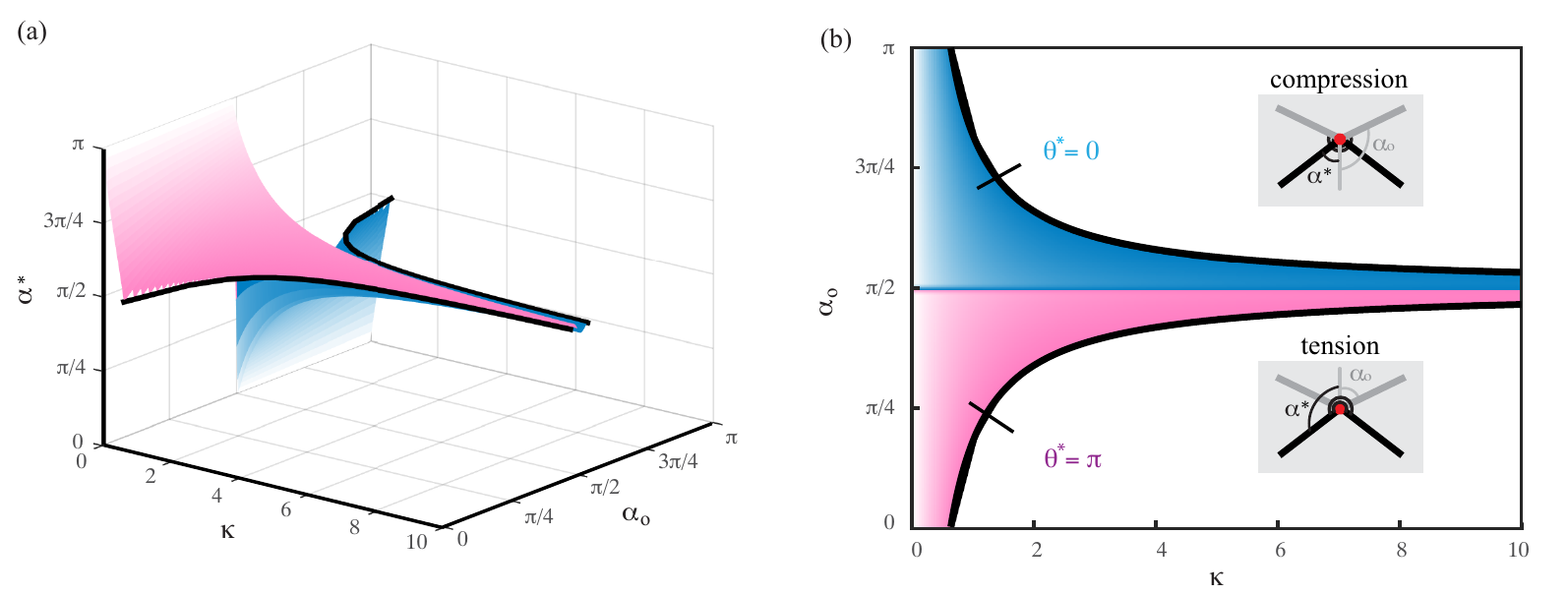}
\caption{\footnotesize Insectile wings: (a) Stable flapping angle $\alpha^*$ as a function of the spring coefficient $\kappa$ and rest angle $\alpha_o$. (b) Projection  onto the $(\kappa,\alpha_o)$ plane, showing the two stable regions for $\theta^*=0$ and $\theta^*=\pi$, separated by $\alpha_o$ = $\pi/2$. 
}
\label{fig:elasticphase}
\end{figure*}

\section{Results}
\label{sec:results}

We examine the behavior of the coupled insectile wings in Eq.~\eqref{eq:eom} for finite spring stiffness $\kappa$ and rest angle $\alpha_o$. Given the left-right symmetry of the wings, it suffices to examine the dynamical behavior of  $\alpha$ and $\theta$ in the range  $[0, \pi]$. In this range,  the equilibrium points $(\alpha^*, \theta^*)$ of Eq.~\eqref{eq:eom} can be identified as a function of $\kappa$ and $\alpha_o$ as follows
\begin{equation}
\begin{split}
\theta^*&=0, \qquad \sin\alpha^* =-\kappa(\alpha^*-\alpha_o), \\
\theta^*&=\pi, \qquad \sin\alpha^*=\kappa(\alpha^*-\alpha_o), \\
\alpha^*&=\pi/2, \ \ \ \cos\theta^*=\kappa(\alpha_o-\pi/2).
\label{eq:equilibria}
\end{split}
\end{equation}
We study the linear stability of these equilibrium points by linearizing Eq.~\eqref{eq:eom} about $(\alpha^*, \theta^*)$, 
\begin{equation}
\begin{split}
\left(\begin{array}{l} \ddot{{\theta}} \\ \ddot{ \alpha}  \end{array} \right) =  
\left(\begin{array}{cc}
-\cos\alpha^*\cos\theta^* & \sin\alpha^*\sin\theta^*  \\
\sin\alpha^*\sin\theta^* &  -\cos\alpha^*\cos\theta^*-\kappa
\end{array} \right)
\left(\begin{array}{l} {\theta} \\ {\alpha}  \end{array} \right)
\label{eq:eomlinear}
\end{split}
\end{equation}
with the understanding that $\theta$ and $\alpha$ in Eq.~\eqref{eq:eomlinear} are small. Linear stability analysis of the equilibrium points in~Eq.~\eqref{eq:equilibria} reveal the existence of two sets of linearly stable points,
\begin{equation}
\begin{split}
\textrm{For } \frac{\pi}{2} \le \alpha_o \le \frac{\pi}{2}+\frac{1}{\kappa}, \qquad  \theta^*&=0 \textrm{ and } \alpha^* \in[0,\dfrac{\pi}{2}] , \\
\textrm{For }\frac{\pi}{2}-\frac{1}{\kappa} \le \alpha_o \le \frac{\pi}{2}, \qquad  \theta^*&=\pi \textrm{ and } \alpha^* \in[\dfrac{\pi}{2},\pi] .
\label{eq:stability}
\end{split}
\end{equation}
Fig.~\ref{fig:elasticphase}(a) illustrates the stable flapping angle $\alpha^*$ as a function of $\kappa$ and $\alpha_o$. Blue and red colors represent acute ($\alpha^*<\pi/2$) and obtuse ($\alpha^*>\pi/2$) flapping angles, respectively. A projection onto the $(\kappa,\alpha_o)$ parameter space is  shown in Fig.~\ref{fig:elasticphase}(b). 
Insets show the physical configurations of the wing system in these equilibrium positions. While the two configurations are the same, the internal torques used to balance the gravitational effects have opposite signs, namely, $\sin\alpha^*=\mp\kappa(\alpha^*-\alpha_o)$, where the minus sign corresponds to $\theta^\ast = 0$, meaning that the internal spring is in a state of compression.

\begin{figure*}
\centering
\includegraphics[width=\linewidth]{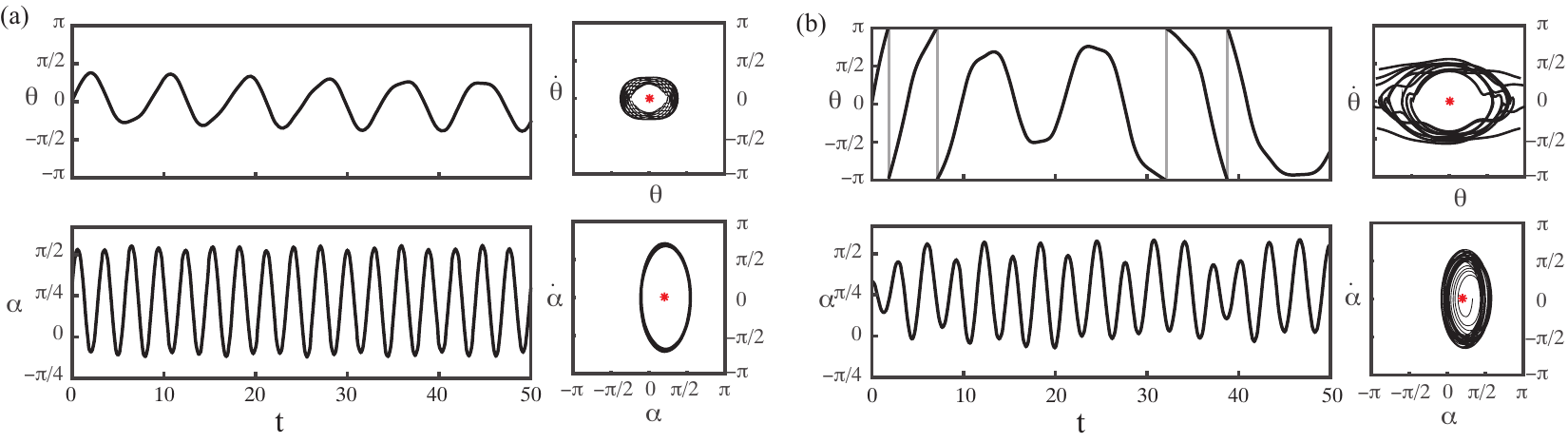}
\caption{\footnotesize Periodic (a) and chaotic (b) behavior around the equilibrium point $(0,\alpha^\ast)$ shown as a red dot for $E=2$, $\kappa=4$ and $\alpha_o=\pi/4$.}
\label{fig:casestudy}
\end{figure*}

An examination of the nonlinear response of the system around one of these equilibrium points shows that it exhibits regular or chaotic behavior depending on initial conditions $(\theta,\alpha,\dot{\theta}, \dot{\alpha})$. Fig.~\ref{fig:casestudy} shows examples of such behavior for $E = 2$,  $\kappa=4$ and $\alpha_o=\pi/4$, and two sets of initial conditions, namely, $(\pi/36,\pi/3,0.88,2.5)$ and $(\pi/36,\pi/3,2.66,0)$.  To characterize the global nonlinear dynamics of the coupled wings, we  construct Poincar\'{e} sections (see, e.g., \cite{Holmes1983}) at $\theta = 0$, which reduces the phase space to a three-dimensional space $(\alpha,\dot{\theta},\dot{\alpha})$. Conservation of energy restricts the dynamical response of the system to lie on a two-dimensional energy surface within this space defined by $\frac{1}{2}(\dot{\alpha}^2+\dot{\theta}^2)-\cos\alpha+\frac{1}{2}\kappa(\alpha-\alpha_o)^2 - E = 0$. 

\begin{figure*}
\centering
\includegraphics[width=\linewidth]{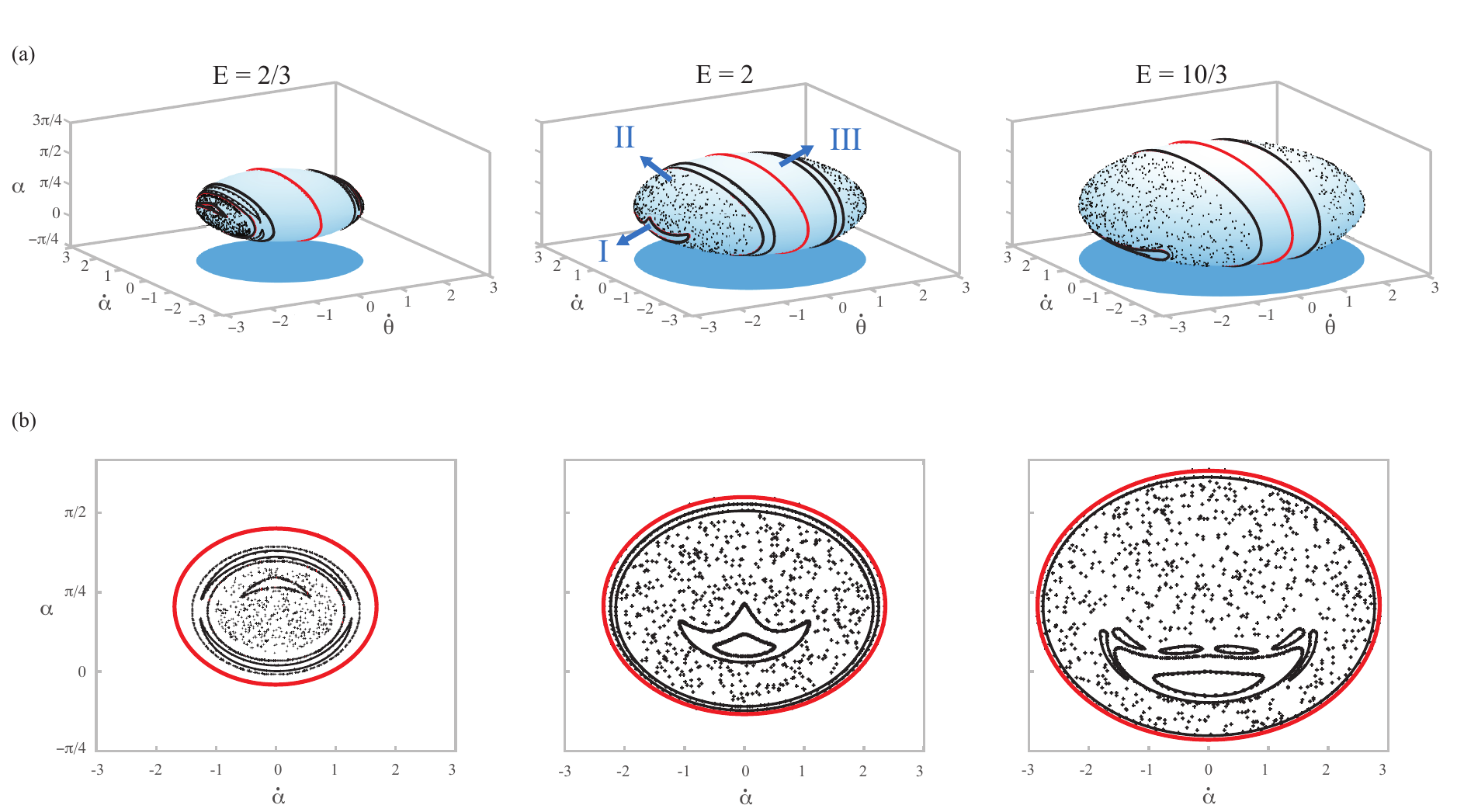}
\caption{\footnotesize Poincar\'{e} sections for $\kappa=4$, $\alpha_o=\pi/4$ and $E = 2/3, 2, 10/3$.} 
\label{fig:poincareE}
\end{figure*}

Fig.~\ref{fig:poincareE}(a) depicts \poincare sections for $\kappa=4$ and three energy levels $E = 2/3, 2, 10/3$. On each energy level, three regions can be distinguished: region I is characterized by periodic rotations; region II shows chaotic behavior; and, region III is characterized by periodic oscillations, which correspond to regular flapping behavior of the wings. Note that as energy increases, regions I and II grow, while region III shrinks. 

\begin{figure*}
\centering
\includegraphics[width=\linewidth]{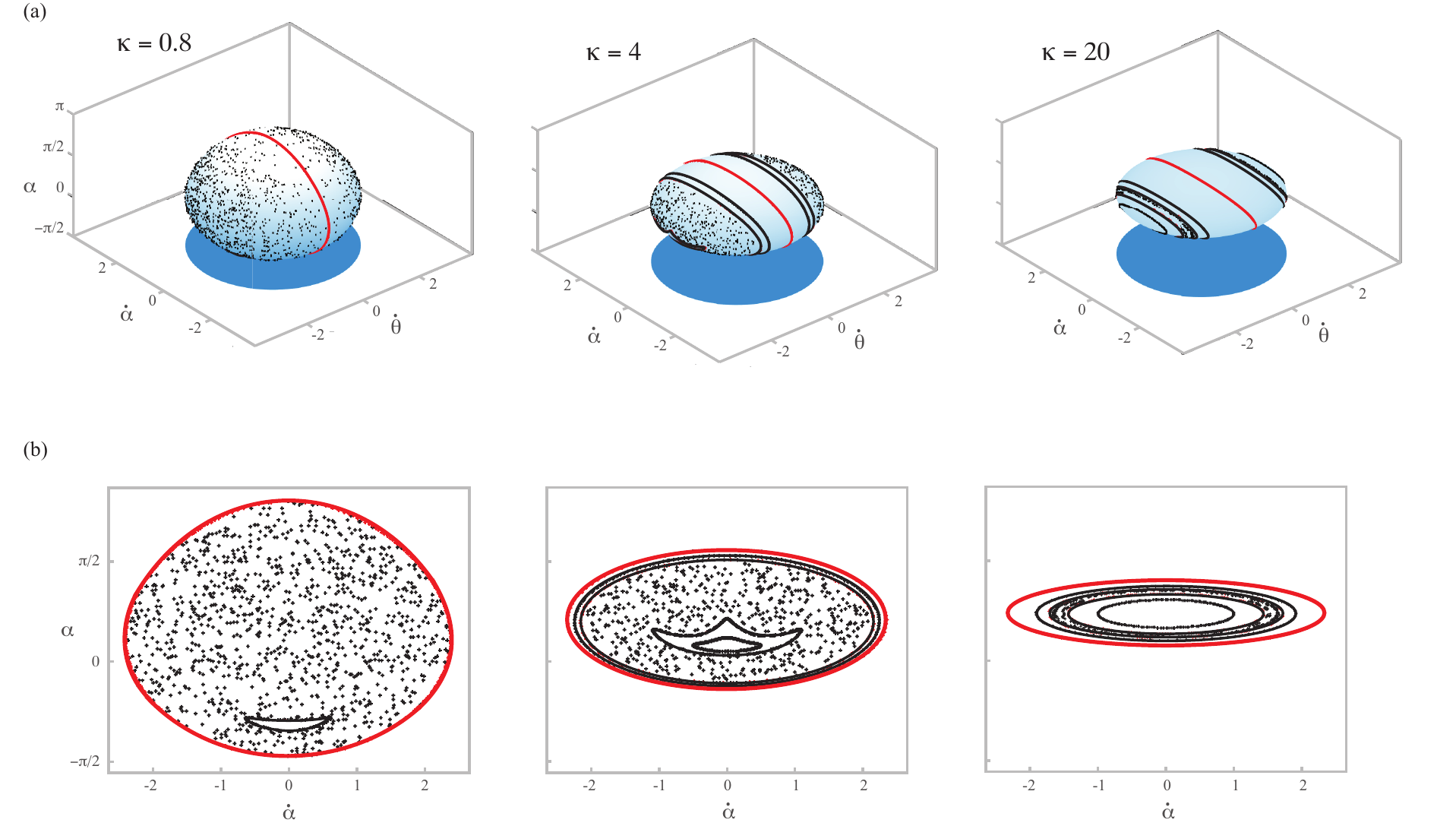}
\caption{\footnotesize Poincar\'{e} sections for $E=2$, $\alpha_o=\pi/4$ and $\kappa = 0.8, 4, 20$. The  case of $\kappa=4$ is same as the one shown in Fig.~\ref{fig:poincareE} ($E=2$).}
\label{fig:poincareK}
\end{figure*}

We now explore the effects of spring stiffness $\kappa$ on the nonlinear response of the wings. Fig.~\ref{fig:poincareK} depicts the energy surfaces for a constant energy value $E=2$ and three different values of spring stiffness, $\kappa= 0.8,4$ and $20$. For small $\kappa$, the behavior of the wings is predominantly chaotic. As $\kappa$ increases, the energy surface turns oblate in the $\alpha$-direction, indicating smaller flapping amplitudes, and the chaotic region II shrinks. A few comments on the limits as $\kappa$ goes to $0$ and $\infty$ are in order here. At $\kappa=0$, the dynamical response is regular as depicted in Fig.~\ref{fig:zerostiff}, but for small non-zero $\kappa$, the behavior becomes immediately irregular. That is to say, weak elastic coupling between the wings induces chaotic flapping. The limit as $\kappa \to 0$ is singular. Stronger elastic coupling between the wings tends to suppress the chaotic behavior, in favor of regular flapping oscillations which decrease as $\kappa \to \infty$. In this non-singular limit, region I expands and shifts towards the rest angle, indicating small periodic flapping about the spring rest position $\alpha_o$. At $\kappa = \infty$, this periodic flapping is suppressed and the wings behave as a rigid pendulum, as mentioned in Section~\ref{sec:model}.

\section{Discussion}
\label{sec:disc}

We proposed an idealized two-dimensional model of insect-like wings. The model consists of two rigid wings connected at their base via an elastic torsional spring. We studied the passive dynamics of this system under gravitational effects, emphasizing the coupling between the wings flapping and rotational motions. We identified regimes where the wings flap stably and others where the wings rotate (roll end over end) or behave chaotically. For a given energy level, the flapping region increases as the stiffness of the torsional spring increases. These findings are qualitatively consistent with insect wings. Insect wings and the flight muscles at their base are known to be very stiff; even at the small operating strains, they can store elastically much of the inertial energy of the oscillating wings \cite{Ellington1985}. More importantly, our results suggest that, by manipulating the stiffness of their flight muscle, insects can maintain periodic flapping when operating at a range of energy levels. It is known that muscle force can be modulated using a number of mechanisms such as changing the muscle stiffness or length (introducing pre-tensioning in the muscle) \cite{Feldman2009}. We conjecture that modulation of muscle stiffness helps insect wings operate
at a range of aerodynamic loads.

\begin{figure*}
\centering
\includegraphics[width=\linewidth]{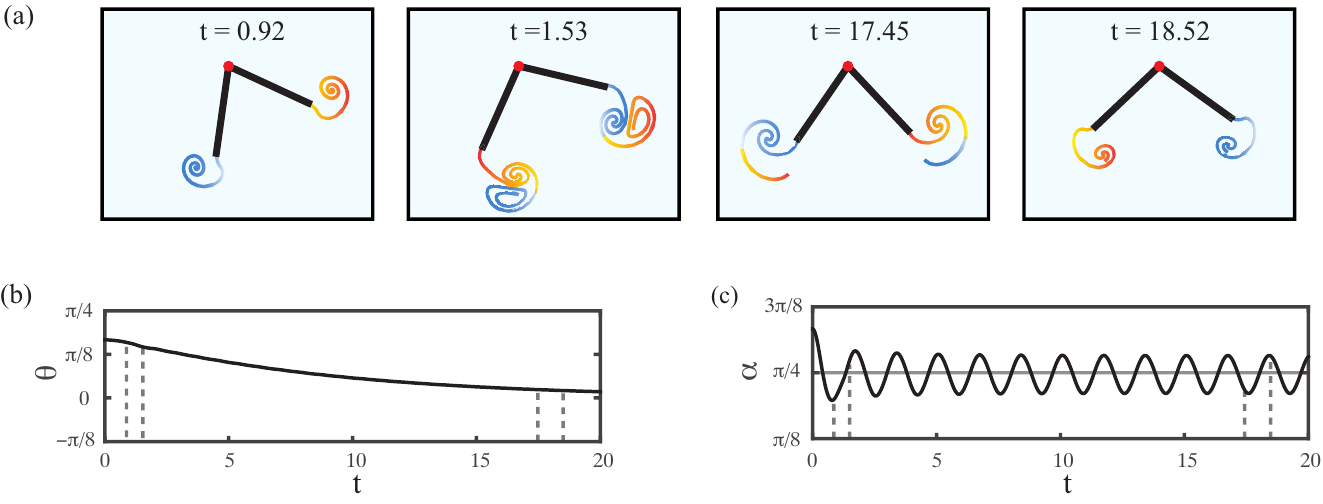}
\caption{\footnotesize  {Elastic wings of spring stiffness $\kappa=80$ and rest angle $\alpha_o=\pi/4$ in rest fluid.   (a) Snapshots of vortex shedding at four time instants highlighted in (b) and (c). (b) Time evolution of the orientation which approaches the vertically down position $\theta=0$. (c) Flapping response  $\alpha$ versus time. Initial conditions are set as $\theta=\pi/6, \alpha = \pi/3, \dot{\theta}=0$ and $\dot{\alpha}=0$. Dissipation time $T_{\rm diss}=1.22$.}
}
\label{fig:influid}
\end{figure*}

The effect of aerodynamic forces on the wings dynamics can be accounted for using a vortex sheet model in the inviscid fluid context, as shown in Fig.~\ref{fig:influid}. Here, the wings are modeled as a bound vortex sheet that satisfies zero normal flow. A point vortex is released at each time step from the two outer edges, and the shed vorticity is modeled as a regularized free sheet \cite{Krasny1986,Nitsche1994,Jones2003,Jones2005,Alben2008,Michelin2008}. No separation is allowed at the apex. We followed the algorithm in \cite{Nitsche1994} for imposing the Kutta condition that determines the amount of circulation shed from the outer two edges at each time step. The vortex sheet model depends on the regularization parameter for the free sheet, which we set to $\delta/l = 0.1$. To emulate the effect of viscosity, we allowed the shed vortex sheet to decay gradually by dissipating each incremental point vortex after a finite time $T_{{\rm diss}}$ from the time it is shed in the fluid. This computational scheme is validated and used in~\cite{Huang2014}. Fig.~\ref{fig:influid} shows a depiction of the wings' flapping behavior, under both gravitational and aerodynamics forces and torques.  Interestingly, the wings flap stably about the vertical position,  even though the initial orientation $\theta$ of the wings is perturbed away from the vertical position. Note that this stable flapping is observed at a larger value of spring stiffness than the ones explored in Fig.~\ref{fig:poincareK}.  A larger stiffness is needed to support the aerodynamic forces and torques due to the surrounding fluid. A detailed study of the aerodynamic effects on the wing behavior will be the subject of a future study.


We conclude by noting that, in addition to neglecting aerodynamic effects, we made a number of assumptions in this paper to ensure the problem is tractable analytically and computationally. We assumed the motion is planar, but three-dimensional wing rotations play an important role in force production in insect wings {\cite{Ellington1996,Dickinson1999,Sun2014}}. We ``lumped" all elastic components into a single torsional spring that couples the dynamics of the two wings. The elasticity of insect flight systems is distributed along the wings and in the flight and thoracic muscles \cite{Pringle2003,Combes2003I,Combes2003II,Wootton1992}. Future extensions of this work will interrogate the effect of each of these simplifying assumptions on the wing flapping dynamics.


%
%

%
\bibliographystyle{ieeetr}
 \bibliography{HuangKanso2015}

%

\end{document}